\documentclass[11pt,epsf,amssymb,ulem,qsymbols]{article}
\usepackage{tabularx}
\usepackage{array}
\usepackage{graphics}
\usepackage{graphicx}
\usepackage{psfrag}
\usepackage{epsfig}
\usepackage{amsmath}
\usepackage{amssymb}
\usepackage{ulem}
\usepackage{setspace}
\usepackage{rotating}
\usepackage{colortbl}
\usepackage{tabularx}
\usepackage{longtable}
\usepackage{multirow}
\makeatletter

\usepackage{verbatim}

\setlongtables

\newcommand{\msbar}{{\overline{\rm MS}}}

\newcommand{\bea}{\begin{eqnarray}}
\newcommand{\eea}{\end{eqnarray}}
\newcommand{\beq}{\begin{equation}}
\newcommand{\eeq}{\end{equation}}
\newcommand{\ec}{\end{center}}
\newcommand{\bc}{\begin{center}}
\newcommand{\gev}{{\rm GeV}}

\newcommand{\pdir}{p\kern -5.2pt\raise 0.2ex\hbox {/}}

\newcommand{\vdir}{v\kern -5.75pt\raise 0.15ex\hbox {/}}
\newcommand{\kdir}{k\kern -5.75pt\raise 0.15ex\hbox {/}}
\newcommand{\epsdir}{\epsilon\kern -5.0pt\raise 0.15ex\hbox {/}}
\newcommand{\bvdir}{\bar{v}\kern -5.75pt\raise 0.15ex\hbox {/}}
\newcommand{\Ddir}{D\kern -7.75pt\raise 0.20ex\hbox {/}}
\newcommand{\Adir}{A\kern -7.75pt\raise 0.20ex\hbox {/}}
\newcommand{\ldir}{l\kern -5.0pt\raise 0.2ex\hbox{/}}
\newcommand{\varepsdir}{\varepsilon\kern -5.5pt\raise 0.15ex\hbox{/}}

\newcommand{\nn}{\nonumber}
\makeatother

\begin{document}
\begin{flushright}
\begin{tabular}{l}
{\small \tt LPT 12-62}\\
{\small \tt LAL 12-219}\\
\end{tabular}
\end{flushright}
\begin{center}
\vskip .8cm\par
{\par\centering \textbf{\LARGE  
\Large \bf $\bar B\to D\tau\bar \nu_\tau$ vs. $\bar B\to D\mu\bar \nu_\mu$ }}\\
\vskip .8cm\par
{\scalebox{.79}{\par\centering \Large  
\sc Damir~Be\v{c}irevi\'c$^a$, Nejc Ko\v{s}nik$^b$ and Andrey Tayduganov$^a$}
{\par\centering \vskip 0.5 cm\par}
{\sl \small 
$^a$~Laboratoire de Physique Th\'eorique (B\^at.~210)~\footnote{Laboratoire de Physique Th\'eorique est une unit\'e mixte de recherche du CNRS, UMR 8627.}\\
Universit\'e Paris Sud, F-91405 Orsay-Cedex, France.\\
$^b$~Laboratoire de l'Acc\'el\'erateur Lin\'eaire, Centre d'Orsay, Universit\'e de Paris-Sud XI,\\
 B.P. 34,  B\^atiment 200, 91898 Orsay Cedex, France, and \\
J. Stefan Institute, Jamova 39, P. O. Box 3000, 1001 Ljubljana, Slovenia }}
\end{center}

\vskip 0.25cm
\begin{abstract}
Recent experimental results for the ratio of the branching fractions of $\bar B\to D^{(\ast)}\tau\bar \nu_\tau$ and $\bar B\to D^{(\ast)}\mu\bar \nu_\mu$ decays came as a surprise and 
lead to a discussion of possibility to constraining New Physics through these modes. Here we focus on ${\cal B}(\bar B\to D\tau\bar \nu_\tau)/{\cal B}(\bar B\to D\mu\bar \nu_\mu)$ 
and argue that the result is consistent with the Standard Model within $2\sigma$, and that the test of compatibility of this ratio with the Standard Model can be done experimentally with a minimal theory input. 
We also show that these two decay channels can provide us with quite good constraints of the New Physics couplings.
\end{abstract}
\vskip 1.5cm

%\setcounter{footnote}{0}
%%%%%%%%%%%  Section 1
\noindent {\bf 1. Introduction:}
Recent experimental result by BaBar~\cite{babar:2012xj} 
\bea\label{eq:1}
R(D)={{\cal B}(\bar B \to D\tau\bar \nu_\tau) \over {\cal B}(\bar B \to D\mu\bar \nu_\mu) }=0.440\pm 0.058\pm 0.042\,,
\eea
seems to indicate a disagreement with the Standard Model (SM) prediction.  In addition to the above ratio, the experimenters also measured the corresponding decays to the final vector meson. Since the latter involve more form factors and can also be experimentally more challenging due to the need of discerning the soft pion events in $B\to D^\ast (\to D\pi)\ell\nu$ from those in $B\to D^{\ast\ast} (\to D\pi)\ell\nu$, we prefer to focus on $\bar B\to D\ell \bar \nu$ decays, even if the issue of properly  handling the soft photon emission in this decay still remains to be solved~\cite{withnejc}.~\footnote{$\ell$ labels the lepton flavor. In practice, $\ell =\mu$ refers to the combined  $\bar B\to D e \bar \nu_e$ and  $\bar B\to D\mu\bar \nu_\mu$, as both leptons can be considered as massless with respect to the heavy mesons involved in the process.} 

To begin with, let us write the relevant hadronic matrix element in the form in which it is usually done in QCD, namely,
\bea\label{eq:1a}
\langle D(p^\prime) \vert \bar c \gamma_\mu b \vert\bar B(p)\rangle  &=&  
\left( p_\mu+p^\prime_\mu - {m_B^2 - m_D^2 \over q^2 } q_\mu \right) F_+(q^2)
+ {m_B^2 - m_D^2 \over q^2 } q_\mu 
 F_0(q^2)  \;,
\eea
that leads to the differential decay rate,   
\begin{align}\label{eq:2}
{d{\cal B}(\bar B \to D\ell \bar \nu_\ell)\over dq^2}& = \tau_{B^0}{G_F^2\vert V_{cb}\vert^2\over 192 \pi^3 m_B^3} \biggl[ c_+^\ell (q^2) \vert F_+(q^2)\vert^2 +  c_0^\ell  (q^2) \vert F_0(q^2)\vert^2\biggr]\nn\\
&= \vert V_{cb}\vert^2 {\cal B}_0  \vert F_+(q^2)\vert^2 \left[ c_+^\ell (q^2)  + 
  c_0^\ell (q^2) \left| { F_0(q^2)\over   F_+(q^2)}\right|^2 \right]\,,
\end{align}
where
\bea\label{eq:cs}
&&c_+^\ell (q^2) =\lambda^{3/2}(q^2,m_B^2,m_D^2) \left[ 1- \frac{3}{2}\frac{ m_\ell^2}{q^2} +  \frac{1}{2}\left(\frac{ m_\ell^2}{q^2}\right)^3\right]\,,\nn\\
&&c_0^\ell (q^2) =m_\ell^2 \ \lambda^{1/2}(q^2,m_B^2,m_D^2) \frac{3 }{2} \frac{ m_B^4}{q^2}\left(1-\frac{ m_\ell^2}{q^2}\right)^2\left(1-\frac{ m_D^2}{m_B^2}\right)^2\,,\nn\\
&&\hfill \nn\\
&& \lambda(q^2,m_B^2,m_D^2) =[q^2 - (m_B+m_D)^2] [q^2 - (m_B-m_D)^2] \,.
\eea
From the above expressions it is obvious that for the massless lepton in the final state the scalar form factor $F_0(q^2)$ does not contribute to the differential branching fraction. 
In the case of $\tau$-lepton, instead, the last term becomes more important and one can question whether or not the coupling to a scalar non-SM particle can be probed through this decay.  Usual assumption is that a charged Higgs boson might give a non-zero contribution in the $b \to c H^-\to c \ell\bar \nu_\ell$ transition~\cite{fred-jernej,stephanie}, a scenario that was recently challenged by the experimental results~\cite{babar:2012xj}. A couple of alternatives have already been proposed~\cite{alternatives1,alternatives2}, and suggestions for further experimental analyses indicated~\cite{others}.

Up to now all the experimental analyses of $B\to D^{(\ast)}$ decays have been made by heavily relying on heavy quark effective theory (HQET)~\cite{HQET}. While HQET provides us with an extremely useful tool in understanding and simplifying the non-perturbative dynamics of QCD in the processes involving heavy-light mesons, at the level of precision aimed at the $B$-factories, the HQET description of the $B\to D^{(\ast)}$ transition matrix element is not as helpful anymore and one should attempt doing the QCD analysis instead. 

In the HQET description of this decay, after taking both meson masses to infinity (or $m_{c,b}\to \infty$), the vector form factor $F_+(q^2)$ --by virtue of the heavy quark flavor symmetry-- is related to the elastic vector form factor and therefore normalized to one at the zero-recoil.~\footnote{The term {\sl vector} ({\sl scalar}) form factor for $F_+(q^2)$ ($ F_0(q^2)$) is related to the fact that in the $t$-channel it couples to the states with quantum numbers $J^P=1^-$ ($0^+$).}  To relate that normalization to the measured branching fraction one needs to make hazardous computation to match HQET with QCD, and include the power corrections that might be uncomfortably large, especially those ${\cal O}(1/m_c^n)$. Worse even, the symmetry point at which the normalization of the form factor is fixed to unity is $q^2_{\rm max}=(m_B-m_D)^2$ [or equivalently $w=p_B\cdot p_D/(m_B m_D)=1$], where there is no phase space, $c_{+,0}^\ell (q^2_{\rm max})=0$, and therefore the assumptions on the shape of the form factor become essential. At first it was believed that the slope of the form factor was enough, but later it became clear that an information about its curvature was indispensable~\cite{CLN,LeYaouanc:2003rn}.  The importance of that issue is obvious since the available phase space rapidly grows with $w$ (for lower $q^2$'s). Clearly, a description of this decay that does not rely on HQET is welcome. That statement should not be viewed as if HQET is not useful any more. It is still the most valuable framework for understanding the dynamics of heavy-light mesons in $B\to D^{\ast\ast}$ semileptonic and non-leptonic decays, and in many other processes, but it is not practical for the exclusive $B\to D^{(\ast)}$ semileptonic modes that are likely to lead to the precision determination of $\vert V_{cb}\vert$, i.e. a determination that requires the least number of assumptions about the underlying QCD dynamics.

\vskip 1.3cm
\noindent {\bf 2. $B\to D\ell \nu$ with minimal theory input:} Let us return to eq.~(\ref{eq:1a}) and note that the range of $q^2$'s available from this decay is large:
\bea
m_\ell^2 \leq q^2 \leq (m_B-m_D)^2=11.63~\gev^2\,.
\eea
\begin{figure}
\hspace*{-.6cm}\includegraphics[width=0.5\textwidth]{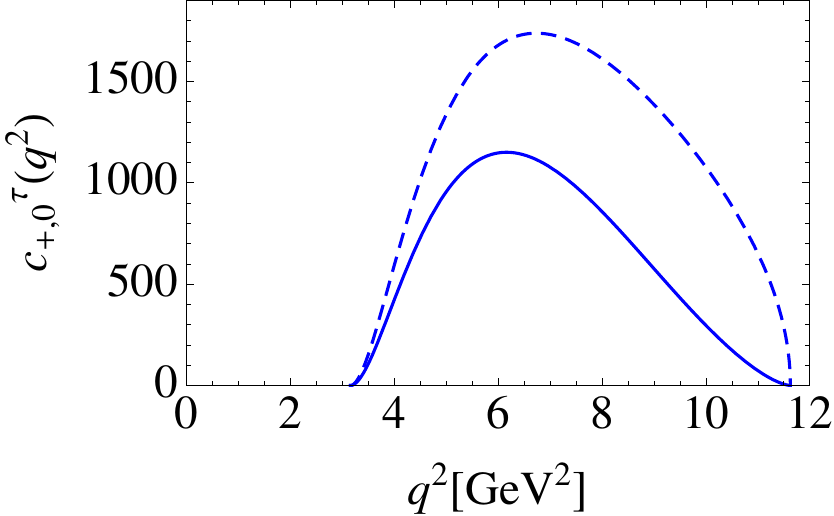}\includegraphics[width=0.53\textwidth]{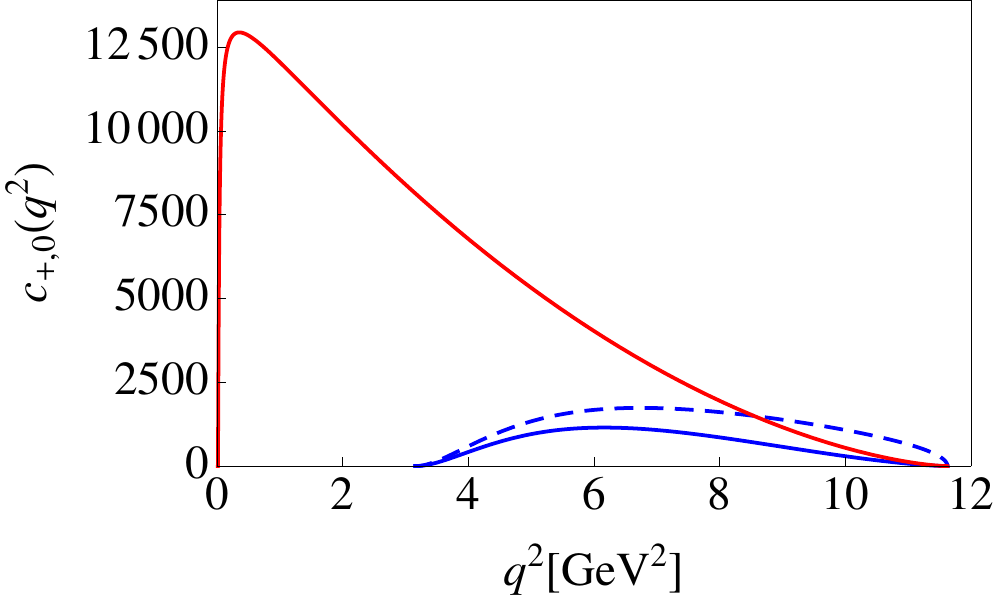}
\caption{\footnotesize  Phase space available from $\bar B\to D\tau\bar \nu_\tau$ is shown in the left plot. Dashed curve corresponds to $c_0^\tau(q^2)$, while the solid curve corresponds to $c_+^\tau(q^2)$. The expressions for $c_{0,+}^\tau(q^2)$ are given in eq.~(\ref{eq:cs}). In the right plot the same  $c_{+,0}^\tau(q^2)$ are plotted together with  $c_+^\mu(q^2)$ (red curve), the phase space available in the case of  massless muon in the final state.}
	\label{fig:1}
\end{figure}
The form factors $F_{0,+}(q^2)$ can be computed on the lattice. The strategy that requires minimum assumptions and allows the precision determination exists and it has been implemented in 
the quenched approximation ($N_{\rm f}=0$)~\cite{nazario}.~\footnote{Another strategy has been proposed and implemented recently in the computation of the $B$-meson decay constants $f_B$ but not in the computation of $B\to D$ transition form factors~\cite{Dimopoulos:2011gx}.} The unquenched results with $N_{\rm f}=2+1$ dynamical light flavors have been recently reported too~\cite{Bailey:2012rr}.
An important constraint on the form factors is that they are equal at $q^2=0$,
\bea
F_+(0) = F_0(0)\,,
\eea
and everywhere else the vector form factor is larger than the scalar one, $\vert F_+(q^2)\vert \geq \vert F_0(q^2)\vert$. This constraint is useful because the scalar form factor is enhanced by $c_0^\tau(q^2)$, c.f. fig.~\ref{fig:1}, and the contributions to the decay rate coming from the vector and scalar form factors are competitive in size.
After taking $m_\mu^2=0$, we can write
\begin{align}
c_0^\mu (q^2) &= 0\,,\nn\\
c_+^\tau (q^2) &= c_+^\mu (q^2) + \Delta c_+(q^2) =  c_+^\mu (q^2) - \lambda^{3/2}(q^2,m_B^2,m_D^2)\  \frac{ m_\tau^2}{2 q^2}  \left[ 3 -  \left(\frac{ m_\tau^2}{q^2}\right)^2\right]\,,
\end{align}
and 
\begin{align}
{\cal B}(\bar B \to D\mu\bar \nu_\mu)&=    \vert V_{cb}\vert^2 {\cal B}_0 \int_{m_\mu^2}^{q_{\rm max}^2}c_+^\mu (q^2)   \vert F_+(q^2)\vert^2 dq^2 \nn\\
&=
 \vert V_{cb}\vert^2 {\cal B}_0 \int_{m_\mu^2}^{m_\tau^2}c_+^\mu (q^2)   \vert F_+(q^2)\vert^2 dq^2 +  \vert V_{cb}\vert^2 {\cal B}_0 \int_{m_\tau^2}^{q_{\rm max}^2}c_+^\mu (q^2)   \vert F_+(q^2)\vert^2 dq^2 \,,
\end{align}
where $q^2_{\rm max}=(m_B-m_D)^2$. On the other hand
\begin{align}
{\cal B}(\bar B \to D\tau\bar \nu_\tau)=&   \vert V_{cb}\vert^2 {\cal B}_0 \int_{m_\tau^2}^{q_{\rm max}^2}   \vert F_+(q^2)\vert^2  \biggl[ c_+^\tau (q^2) +  c_0^\tau  (q^2) \left| { F_0(q^2)\over   F_+(q^2)}\right|^2 \biggr] dq^2 \nn\\
=&
 \vert V_{cb}\vert^2 {\cal B}_0 \int_{m_\tau^2}^{q_{\rm max}^2} c_+^\mu (q^2)   \vert F_+(q^2)\vert^2 dq^2 \nn\\
 &+
   \vert V_{cb}\vert^2 {\cal B}_0 \int_{m_\tau^2}^{q_{\rm max}^2}   \vert F_+(q^2)\vert^2  \biggl[ \Delta c_+ (q^2) +  c_0^\tau  (q^2) \left| { F_0(q^2)\over   F_+(q^2)}\right|^2 \biggr] dq^2\,,
\end{align}
so that the ratio from eq.~(\ref{eq:1}) can be written as
\bea
R(D)= {1+ R_\tau\over 1+ R_\mu}\,,
\eea
with
\begin{align}\label{r12}
&R_\tau ={\displaystyle{  \int_{m_\tau^2}^{q_{\rm max}^2}   \vert F_+(q^2)\vert^2  \biggl[ \Delta c_+ (q^2) +  c_0^\tau  (q^2) \left| { F_0(q^2)\over   F_+(q^2)}\right|^2 \biggr] dq^2 }\over  \displaystyle{  \int_{m_\tau^2}^{q_{\rm max}^2} c_+^\mu (q^2)   \vert F_+(q^2)\vert^2 dq^2}}\,,\quad
R_\mu ={  \displaystyle{  \int_{m_\mu^2}^{m_\tau^2} c_+^\mu (q^2)   \vert F_+(q^2)\vert^2 dq^2}\over  \displaystyle{  \int_{m_\tau^2}^{q_{\rm max}^2} c_+^\mu (q^2)   \vert F_+(q^2)\vert^2 dq^2}}\,.
\end{align}
Most of these integrals can be evaluated by using the experimentally determined form factor $\vert V_{cb}\vert G(w)$, extracted from the differential branching fraction in ref.~\cite{Aubert:2009ac}, and related to $\vert V_{cb}\vert F_+(q^2)$ via,
\bea
q^2= m_B^2+m_D^2-2m_Bm_Dw \,,\quad F_+(q^2) = \left. {m_B+m_D\over \sqrt{4 m_B m_D}} G(w)\right|_{w(q^2)}.
\eea
For the numerator in $R_\mu$ the lowest three $q^2$ bins, each containing a large fraction of events, lead to an accurate result,
\bea
\vert V_{cb}\vert^2  \int_{m_\mu^2}^{m_\tau^2} c_+^\mu (q^2)   \vert F_+^{\rm exp.}(q^2)\vert^2 dq^2 = 28.7\,.
\eea  
For the denominators in $R_{\mu,\tau}$ we need the form factor $F_+(q^2)$ for $q^2\in [m_\tau^2, q_{\rm max}^2]$, which is difficult to extract from experiment alone because of the smallness of phase space at larger $q^2$'s (smaller $w$), as we can see from fig.~\ref{fig:1}. One can instead combine the experimental results obtained within $q^2\in [m_\tau^2, 8~\gev^2]$, with the lattice QCD results obtained for $q^2\in [8~\gev^2, q_{\rm max}^2]$~\cite{nazario,Bailey:2012rr}, \begin{align}
 \int_{m_\tau^2}^{q_{\rm max}^2} c_+^\mu (q^2)  & \vert F_+(q^2)\vert^2 dq^2 =\nn\\
&   \int_{m_\tau^2}^{8\ \gev^2} c_+^\mu (q^2)   \vert F_+^{\rm exp.}(q^2)\vert^2 dq^2 +  \int^{q_{\rm max}^2}_{8\ \gev^2} c_+^\mu (q^2)   \vert F_+^{\rm latt.}(q^2)\vert^2 dq^2 \,,
\end{align}
which, after using $\vert V_{cb}\vert =0.0411(16)$~\cite{ckm} to multiply the lattice results, leads to
\bea
\vert V_{cb}\vert^2  \int_{m_\tau^2}^{q_{\rm max}^2} c_+^\mu (q^2)   \vert F_+(q^2)\vert^2 dq^2 = (26.3\pm 1.0) + (4.9\pm 0.4) = 31.1\pm 1.1\,.
\eea
Notice that due to the phase space suppression, the last range of $q^2$'s in which we used the lattice data,  makes less than $10\%$ ($20\%$) with respect to the full range of $q^2$ accessible from this decay with muon ($\tau$) in the final state. 
For the numerator in $R_\tau$ the ratio between the scalar and vector form factors is needed too. Lattice QCD results of refs.~\cite{nazario,Bailey:2012rr} can be converted to the form factors employed here by
\begin{align}
&F_+(q^2)= \left.{m_B+m_D\over\sqrt{4m_Bm_D} }h_+(w) - {m_B - m_D\over\sqrt{4m_Bm_D} }h_-(w)\right|_{w(q^2)} ,\nn\\
&F_0(q^2)= \sqrt{m_Bm_D} \left( {w+1\over m_B+m_D}  h_+(w) - {w-1 \over m_B-m_D} h_-(w)\right)_{w(q^2)} \,,
\end{align} 
and after combining them, we see that the ratio $F_0(q^2)/F_+(q^2)$ exhibits a linear $q^2$ behavior, which with the intercept fixed by $F_0(0)/F_+(0)=1$ allows for an accurate determination of the slope $\alpha$,
\bea\label{eq:4}
{F_0(q^2)\over F_+(q^2)} = 1 - \alpha\ q^2\,,
\eea
as shown in fig.~\ref{fig:2}.~\footnote{Please note that the plotted values corresponding to the lattice results with $N_{\rm f}=2+1$ are obtained from $h_\pm (w)$ that we read off from figs.~6 and 7 of ref.~\cite{Bailey:2012rr}. } We obtain $\alpha=0.020(1)~\gev^{-2}$ from the values reported in~\cite{nazario}, and $\alpha=0.022(1)~\gev^{-2}$ from the results of ref.~\cite{Bailey:2012rr}, i.e. slightly smaller than the naive pole dominance model would suggest $\alpha= 1/m_{B_c^\ast}^2 = 0.025~\gev^{-2}$, and very close to the result of  the  model of ref.~\cite{Melikhov:2000yu}, $\alpha = 0.022~\gev^{-2}$, or $\alpha = 0.021(2)~\gev^{-2}$, as obtained in the recent QCD sum rule analyses~\cite{Faller:2008tr,Azizi:2008tt}. 
\begin{figure} 
\hspace*{-.9cm}\includegraphics[width=0.55\textwidth]{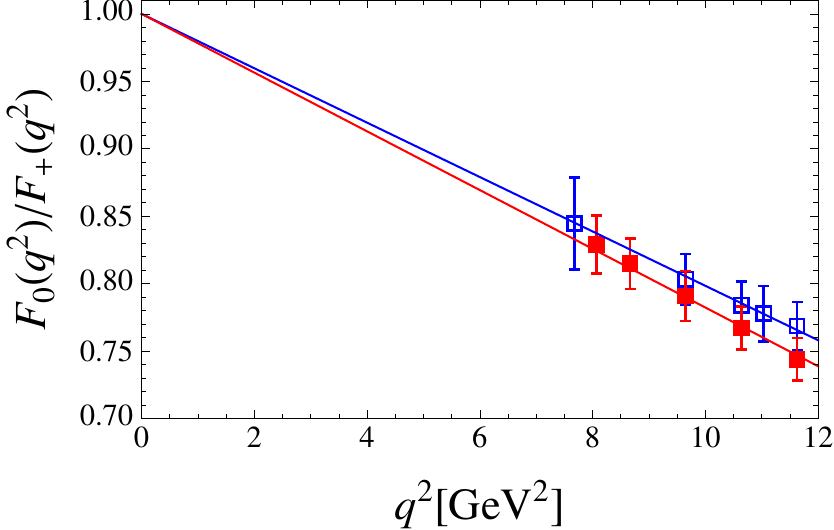}\includegraphics[width=0.55\textwidth]{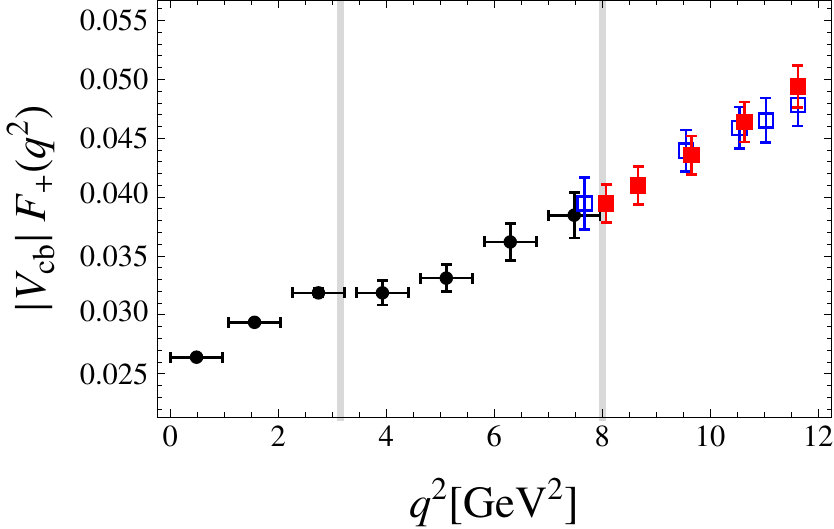}\
\caption{\footnotesize In the left panel we show the ratio of $B\to D$ form factors obtained in the lattice QCD simulations in quenched approximation (empty symbols)~\cite{nazario}, and those in which $N_{\rm f}=2+1$ dynamical flavors are included (filled symbols)~\cite{Bailey:2012rr}.  In the right plot we show the three regions used in eq.~(\ref{r12}). The same lattice data are used in the large $q^2$ region, while the $\vert V_{cb}\vert F_+(q^2)$ at $q^2\lesssim 8~\gev^2$ is extracted from the measured differential branching fraction for $\bar B\to D\mu\bar \nu$~\cite{Aubert:2009ac}.}
	\label{fig:2}
\end{figure}
By using $\alpha= 0.021(1)$, and with the help of eq.~(\ref{eq:4}), for the numerator of $R_\tau$ we obtain $-12.9\pm 0.7$, which finally gives
\bea\label{our-value}
R(D)=0.31\pm 0.02\,,
\eea
which is more than $1$- but less than $2$-$\sigma$ below the BaBar result~(\ref{eq:1}).
\vskip 1.3cm
\noindent {\bf 3. New Physics:} As we saw above the value we obtain is consistent with experiment to within $2\sigma$. Since $R(D)$ requires a minimal non-perturbative QCD theory input, it is tempting to check the constraints on New Physics that one can infer from the comparison between theory and  the BaBar result~(\ref{eq:1}). In a generic New Physics scenario without right-handed neutrino that preserves the lepton flavor universality (LFU), a coupling from this decay to the scalar, vector and tensor operators can be described via~\footnote{A coupling to the pseudoscalar and axial operators cannot be studied in this decay as the corresponding matrix elements vanish due to parity.}
\begin{align}\label{eq:lagr2}
{\cal H}_{\rm eff}=  -{ \sqrt{2} G_F V_{cb}}&\left[( \bar c \gamma_\mu b)( \bar \ell_L \gamma^\mu\nu_L)  + g_V (\bar c \gamma_\mu b)( \bar \ell_L \gamma^\mu\nu_L)  \right.\nn\\
&\left. + g_S(\mu)( \bar c b )(\bar \ell_R \nu_L) 
+ g_T(\mu)( \bar c \sigma_{\mu \nu}b )(\bar \ell_R \sigma^{\mu \nu} \nu_L)\right] + {\rm h.c.} \,,
\end{align}
where the dimensionless couplings $ g_{V,S,T}\propto m_W^2/m_{\rm NP}^2$, with $m_{\rm NP}$ being the New Physics scale. The differential decay rate from eq.~(\ref{eq:2}) now becomes,
\begin{align}\label{eq:0}
{d{\cal B}(\bar B \to D\ell \bar \nu_\ell)\over dq^2} = \vert V_{cb}\vert^2 {\cal B}_0  \vert F_+&(q^2)\vert^2 \left\{ \vert 1 + g_V\vert^2 c_+^\ell (q^2) + \vert g_T(\mu)\vert^2 c_T^\ell (q^2 )   \left|{F_T(q^2,\mu)\over F_+(q^2)}\right|^2\right. \nn\\
&   + c_{TV}^\ell (q^2 ) \ {\rm Re} \left[(1+g_V) g_T^\ast (\mu)  \ {F_T(q^2,\mu)\over F_+(q^2)} \right] \nn\\
&\left.  +  \left| (1+g_V) -  {q^2\over m_\ell}{g_S(\mu) \over m_b(\mu)-m_c(\mu)} \right|^2c_0^\ell (q^2) \left| { F_0(q^2)\over   F_+(q^2)}\right|^2 \right\}\,,
\end{align}
where
\begin{align}
c_T^\ell (q^2,\mu)&= \lambda^{3/2}(q^2,m_B^2,m_D^2) {2  q^2 \over (m_B+m_D)^2}  \left[1- 3 \left(\frac{ m_\ell^2}{q^2}\right)^2 +2 \left(\frac{ m_\ell^2}{q^2}  \right)^3\right] \,,\nn\\
c_{TV}^\ell (q^2,\mu) &=  {6 m_\ell \over   m_B+m_D } \  \lambda^{3/2}(q^2,m_B^2,m_D^2) \left(1-  \frac{ m_\ell^2}{q^2} \right)^2\,,
\end{align}
and the form factor $F_T(q^2,\mu)$ is defined as
\bea\label{eq:fTT}
\langle D(p^\prime) \vert \bar c \sigma_{\mu\nu} b \vert\bar B(p)\rangle  = - i
\left( p_\mu p^\prime_\nu -  p^\prime_\mu p_\nu \right) {2 \ F_T(q^2,\mu) \over m_B+m_D} \;.
\eea
The above formulas agree with those reported in ref.~\cite{Kronfeld:2008gu}. A possibility to discern a small $g_V$ from this experiment seems very unlikely. It was recently searched while checking the unitarity of the first raw of the CKM matrix~\cite{Antonelli:2010yf} and was found to be consistent with zero. On the other hand $g_S(\mu)\neq 0$ is plausible but its value is expected to be very small as the left-right operator lifts the helicity suppression and affects both $\bar B\to D\tau \bar  \nu$ and $\bar B\to D\mu \bar \nu$ decays.  Its non-zero value could also be a source of difficulties for the $D^0-\bar D^0$ mixing amplitude, as the left-right operators are enhanced by the factor $m_D^2/m_c^2$ with respect to the left-left (SM) contribution. A sizable effect could also be seen in $D\to V\gamma$ decays, as those too are governed by the loops, with the down-type quarks propagating in the loop and therefore sensitive to $g_S(\mu)\neq 0$~\cite{jernej-gino}. From $R(D)$ alone we get a very loose constraint on $g_S(m_b)$ (c.f. the contour plot in fig.~\ref{fig:3}). Requiring the compatibility of the theoretical expression for ${\cal B}(\bar B\to D\mu\bar \nu_\mu)$ obtained by using eq.~(\ref{eq:0}) with $g_S(m_b)\neq 0$ and $g_V=g_T(m_b)=0$, and the measured value~\cite{Aubert:2009ac},  restricts the allowed $g_S(m_b)$ to a small region also indicated in fig.~\ref{fig:3}. For example, if $g_S(m_b)$ is real then the $1\sigma$ compatibility with experiment would allow $-0.37\leq g_S(m_b) \leq -0.05$, while the $3\sigma$ compatibility would amount to  $-0.53\leq g_S(m_b) \leq +0.20$.
Note that $g_S(m_b)$, extracted from (tree level) semileptonic process should be run to $\mu=m_{\rm NP}$ before using it in the loop induced processes. With the help of the $\msbar$ mass anomalous dimension~\cite{Chetyrkin}, we find $g_S(m_{\rm NP}=1\ {\rm TeV}) = 0.58 \ g_S(m_{b})$. In the above discussion we assumed a common practice of using the standard quark masses renormalized in the $\msbar$ scheme at $\mu=m_b$, the values of which can be found in ref.~\cite{PDG}.

If in eq.~(\ref{eq:0}) we  set $g_V=g_S(m_b)=0$ and allow for $g_T(m_b)\neq 0$, then the possible values for the real and imaginary parts that are compatible with $R(D)$~\cite{babar:2012xj}  are those in the contour plot shown in fig.~\ref{fig:3}. To do that we obviously needed the tensor form factor which has not been computed on the lattice nor in the QCD sum rules. To our knowledge it was only computed in the model of ref.~\cite{Melikhov:2000yu} from which we learn that $F_T(q^2)/F_+(q^2) = 1.03(1)$ is a constant, in agreement with naive expectations based on the pole dominance. Again, $R(D)$ alone is not constraining much the possible values of $g_T(m_b)$, and the compatibility with the measured  ${\cal B}(\bar B\to D\mu\bar \nu_\mu)$~\cite{Aubert:2009ac} helps selecting a smaller region, also shown in fig.~\ref{fig:3}. If ${\rm Im}\ g_T(m_b)=0$, we obtain $0.3\leq g_T(m_b) \leq 1.5$ and $-0.6\leq g_T(m_b) \leq 2.1$, from the requirement of respective $1$- and $3\sigma$ compatibility with both experimental $R(D)$ and ${\cal B}(\bar B\to D\mu\bar \nu_\mu)$. Note again that $
g_T(m_{\rm NP}=1\ {\rm TeV}) = 0.82 \ g_T(m_{b})$, where we used the QCD anomalous dimension of the tensor density operator~\cite{Gracey:2000am}. 
\begin{figure}
\hspace*{-.8cm}\includegraphics[width=0.528\textwidth]{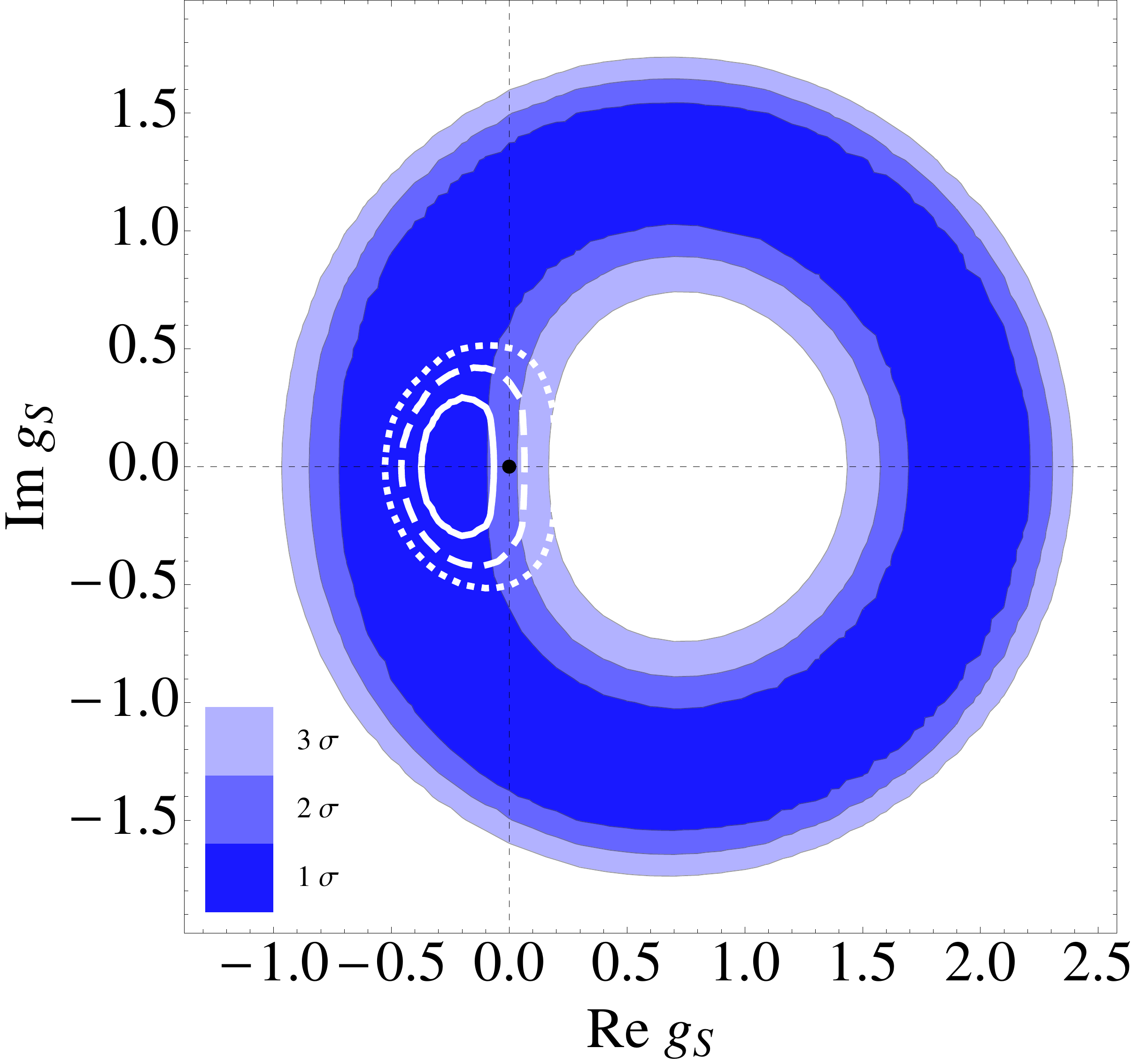}\quad\includegraphics[width=0.5\textwidth]{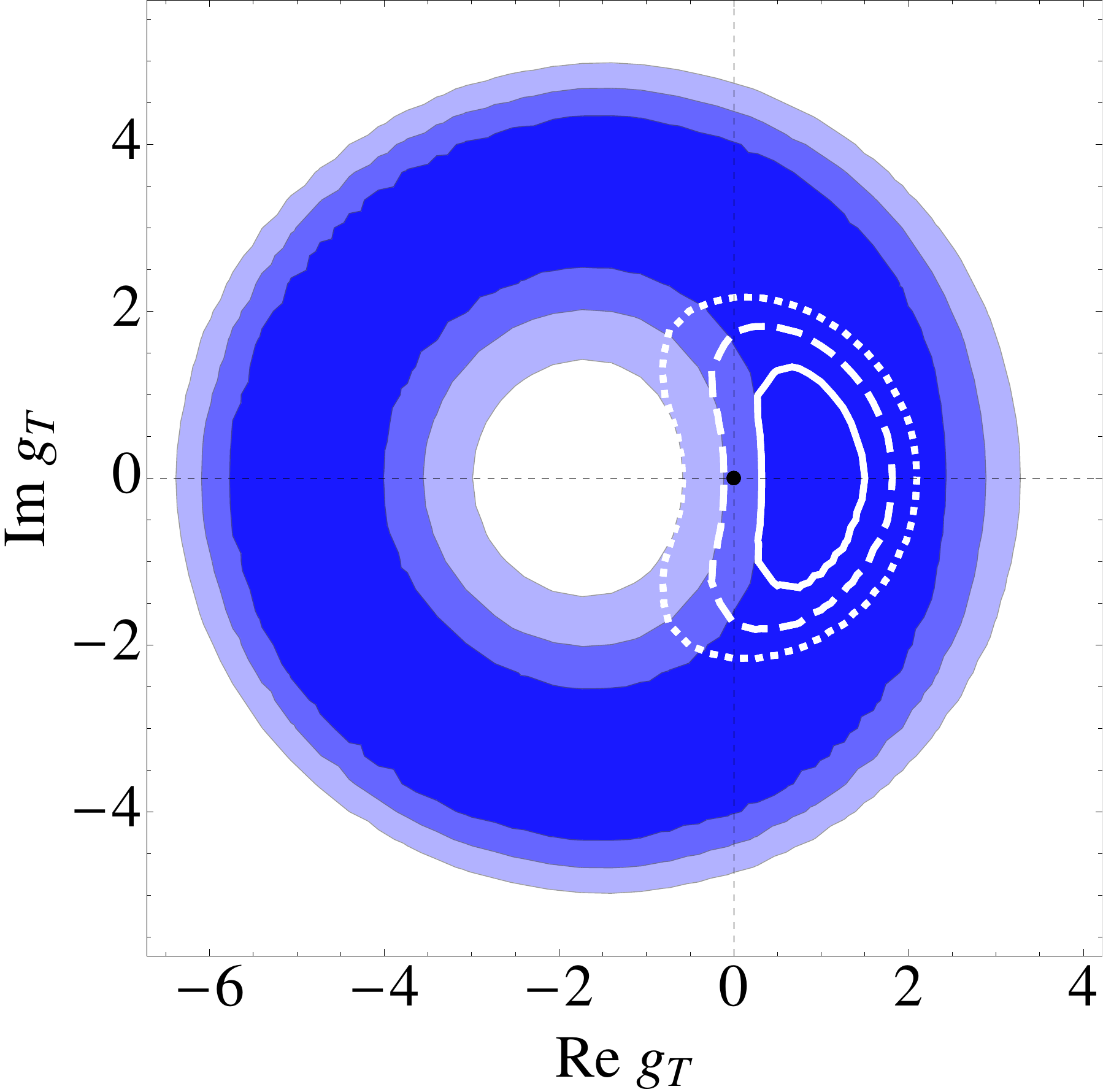}\\
\caption{\footnotesize  Regions of allowed values for $g_S(m_b)$ and $g_T(m_b)$, compatible with experimentally measured $R(D)$. The first (statistical) error in~(\ref{eq:1}) is treated as Gaussian, while the second (systematic) as uniform. The small region within the  solid, dashed and dot-dashed white curves correspond to the respective $1$-, $2$- and $3$-$\sigma$ compatibility with both ${\cal B}^{\rm (exp)}(\bar B\to D\tau\bar \nu_\tau)$ and ${\cal B}^{\rm (exp)}(\bar B\to D\mu\bar \nu_\mu)$. If the LFU is broken the small region disappears and the above contour plot describes $g_{S,T}^\tau (m_b)$. The thick dot represents the Standard Model value, namely $g_{S,T}(m_b)=0$.}
	\label{fig:3}
\end{figure}

As we see from the above discussion, one could easily bridge the gap between experimental result and the SM prediction for ${\cal B}(\bar B \to D \tau \bar \nu_\tau)$ by invoking the New Physics effects that preserve LFU. If we give up LFU,  the couplings $g_{V,S,T}(\mu)$ in the Lagrangian~(\ref{eq:lagr2}) become $g_{V,S,T}^\ell(\mu)$, dependent on the lepton species too. Many concrete New Physics models, such as various types of Two Higgs Doublet Model (2HDM), break LFU and, for example, the scalar coupling $g_S^\ell (\mu)$ becomes proportional to the mass of the charged lepton~\cite{stephanie}. If so, the constraint from ${\cal B}(\bar B \to D \mu \bar \nu_\mu)$ is a factor of $m_\mu/m_\tau$ less sensitive to New Physics than in the case of LFU illustrated above. The net effect would be that the small regions within the white curves depicted in fig.~\ref{fig:3} would simply disappear, and the preferred regions of the scalar coupling $g_S^\tau (m_b)$ would be those represented by the blue contours in the left plot of fig.~\ref{fig:3}.

\vskip 1.3cm
\noindent {\bf 4. Concluding comments:} In closing this letter we would like to make the following comments:
\begin{itemize}
\item Assuming the lepton flavor universality, which has been experimentally verified to a very good accuracy~\cite{PDG}, the result~(\ref{eq:1}) could be an indication of New Physics, if incompatibility with the Standard Model is indeed  shown to be significant (more than $3\sigma$). That test  [of compatibility with the Standard Model] can be made experimentally, with a minimal theory input, as discussed in this letter. 
Here we used the lattice QCD results for $F_+(q^2)$ at larger $q^2$'s because the full branching fractions 
were reported in ref.~\cite{babar:2012xj}.  Our value~(\ref{our-value}) is compatible with experiment within less than $2\sigma$.  If, instead of measuring the full branching fractions 
for both decay modes, the experimenters made a cut at about $q^2\approx 8~\gev^2$, then the full shape of the needed vector form factor could be reconstructed from the differential 
branching fraction of $\bar B\to D\mu\bar \nu$~\cite{fred-jernej,stephanie}. The only information needed from theory is then the ratio of the scalar and vector 
form factors~(\ref{eq:4}), which is quite accurately known from lattice QCD, with the values that agree with quark models, and with recent QCD sum rule studies. 
We hope such an analysis will be made by both BaBar and Belle. By using the vector form factor data from ref.~\cite{Aubert:2009ac} only, and by integrating the decay rates up to $q^2_{\rm cut}=8\ \gev^2$, we obtain 
\bea
\left. {{\cal B}(\bar B \to D\tau\bar \nu_\tau) \over {\cal B}(\bar B \to D\mu\bar \nu_\mu) }\right|_{q^2\leq 8\ \gev^2} =0.20\pm 0.02\,.
\eea
\item $\bar B\to D\mu\bar \nu$ decay can be the mode allowing the most reliable extraction of $\vert V_{cb}\vert$ as it requires the least number of assumptions. A discussion made in ref.~\cite{Aubert:2009ac} showed that one can find a range of $q^2$'s in which both the experimental and lattice QCD errors can be kept small and therefore allow for a very clean extraction of $\vert V_{cb}\vert$.  As for the large measured value for ${\cal B}(\bar B \to D\tau\bar \nu_\tau)$, an independent measurement by Belle is necessary. A measure of its  partial branching fraction would also be helpful to permit a direct comparison with the lattice QCD results for the form factors. In both cases, a special care should be devoted 
to the systematics related to the presence of $B\to D\ell\nu\gamma$ events in the selected sample, with photon being soft.  Such events affect the neutral and the charged $B$-meson semileptonic decays differently~\cite{withnejc}. 
\item As shown above, the measured $R(D)$ and ${\cal B}(\bar B\to D\mu\bar \nu_\mu)$ provide quite good constraints on the new physics couplings $g_{S,T}(m_b)$, that then can be used at the loop induced processes, after running to $g_{S,T}(m_{\rm NP})$. 
\item An experimental  study of $B_s \to D_s \ell \nu$ decay rates would be even more advantageous. It would eliminate the chiral extrapolations of the lattice results in the valence light quark mass. 
A fully unquenched lattice QCD study along the lines presented in ref.~\cite{nazario} would be very welcome too. It would be an important 
independent check of the results of ref.~\cite{Bailey:2012rr}.
\end{itemize}

\vskip 1.6cm
\noindent {\bf Acknowledgment}: It is a pleasure to thank S.~Descotes-Genon, F.~Mescia and the authors of ref.~\cite{alternatives1} for discussions on the topics related to this letter.  We also thank D.~Du for help with the results of ref.~\cite{Bailey:2012rr}, and Z.~Ligeti for pointing out to us the inconsistency of the sign of the form factor in eq.~(\ref{eq:fTT}), as defined in the previous version of this paper, if it was to be kept positive and consistent with the heavy quark limit.

\end{document}